\documentclass[12pt,letterpaper]{article}

\usepackage{cite}
\usepackage{amsmath,amssymb,amsfonts,amsthm}
\usepackage{algorithmic}
\usepackage{graphicx}
\usepackage{textcomp}
\usepackage[margin=1in]{geometry}

\usepackage{cite}
\usepackage{subcaption}

\usepackage{pgfplots}
\usepackage{tikz}
\usepackage{tikzscale}
\usepackage[american resistor, european voltage, european current]{circuitikz}
\usetikzlibrary{arrows,shapes,positioning}
\usetikzlibrary{decorations.markings,decorations.pathmorphing,decorations.pathreplacing}
\usetikzlibrary{calc,patterns,shapes.geometric}
\usetikzlibrary{arrows}
\usetikzlibrary[arrows]
\usetikzlibrary{shapes.misc}

\DeclareMathAlphabet{\pazocal}{OMS}{zplm}{m}{n}

\newtheorem{theorem}{Theorem}[section]

\newtheorem{lemma}{Lemma}[section]

\newtheorem{definition}{Definition}[section]

\begin{document}
\title{	Passivity-based Decentralized Control for Discrete-time Large-scale Systems}
\author{Ahmed Aboudonia, Andrea Martinelli, and John Lygeros}
\date{}

\maketitle
\pagestyle{empty}
\thispagestyle{empty}

\begin{abstract}
Passivity theory has recently contributed to developing decentralized control schemes for large-scale systems. Many decentralized passivity-based control schemes are designed in continuous-time. It is well-known, however, that the passivity properties of continuous-time systems may be lost under discretization. In this work, we present a novel stabilizing decentralized control scheme by ensuring passivity for discrete-time systems directly and thus avoiding the issue of passivity preservation. The controller is synthesized by locally solving a semidefinite program offline for each subsystem in a decentralized fashion. This program comprises local conditions ensuring that the corresponding subsystem is locally passive. Passivity is ensured with respect to a local virtual output which is different from the local actual output. The program also comprises local conditions ensuring that the local passivity of all subsystems implies the asymptotic stability of the whole system. The performance of the proposed controller is evaluated on a case study in DC microgrids.
\end{abstract}


\let\thefootnote\relax\footnote{The manuscript was first submitted in September 2020. This work was supported by the Swiss Innovation Agency Innosuisse under the Swiss Competence Center for Energy Research SCCER FEEB$\&$D and by the European Research Council	(ERC) under the European Union’s Horizon 2020 research and innovation programme grant agreement OCAL, No. 787845. (Corresponding Author: Ahmed Aboudonia) }
\let\thefootnote\relax\footnote{The authors are with the Automatic Control Laboratory, Department of Electrical Engineering and Information Technology, ETH Zurich, 8092 Zurich, Switzerland. Their email addresses are	{\tt\small $\{$ahmedab,andremar,lygeros$\}$@control.ee.ethz.ch}}

\section{Introduction}
\label{sec:introduction}
Passivity theory has proven to be useful for designing feedback controllers for linear and nonlinear systems (e.g. see \cite{kottenstette2014relationships}). Such controllers have been used in many applications such as robotics \cite{albu2007unified} and energy systems \cite{mukherjee2012building}. Various efforts have been also devoted to develop robust \cite{bao2003robust} and adaptive  \cite{wang2012passivity} passivity-based controllers.
Passivity theory has recently also contributed to developing decentralized control schemes for large-scale systems 
\cite{nahata2020passivity}. 
Many passivity-based control schemes are designed in continuous-time. 
It is well-known, however, that the passivity properties of continuous-time systems are lost under discretization due to the resulting energy leakage of the zero-order-hold \cite{stramigioli2002novel}. Hence, various methods are developed in which passivity is preserved under discretization, for example, by using small sampling times \cite{oishi2010passivity} or by introducing virtual outputs \cite{costa2006preserving}. The above methods are mainly developed for centralized systems.

In this paper, we propose a passivity-based decentralized control scheme for a class of large-scale systems which can be decomposed into smaller dynamically-coupled subsystems. Unlike the above-mentioned literature which considers passivating the continuous-time system and then discretizing it while maintaining passivity, we design the proposed controller directly in discrete-time. For each susbsystem, we synthesize a local state-feedback controller which depends on the states of the corresponding subsystem only, resulting in a decentralized architecture. Each local controller is synthesized by locally solving a convex optimization problem independently. 

Each problem comprises conditions to ensure passivity of the corresponding subsystem. Passivity is ensured with respect to a virtual output which is different from the actual output of the subsystem. This virtual output is a combination of the actual outputs of the corresponding subsystem and its neighbours. Besides the control gains, the optimization problem is solved for the storage function, the dissipation rate and the virtual output of the corresponding subsystem. Additional local constraints on the virtual output and the dissipation rate are added to each optimization problem to ensure that the local passivity of all subsystems guarantees the asymptotic stability of the overall system. The efficacy of the proposed controller is demonstrated by implementing it on a DC microgrid model.

One could also consider synthesising decentralised controllers in a centralised way. This would require the information about all dynamics of all subsystems to be available centrally. Our approach obviates this need by also performing the synthesis of the decentralised controller in a decentralised manner. Furthermore, the proposed method does not suffer from the conservative performance associated with decentralized control approaches that treat the coupling terms as bounded disturbances (e.g. see \cite{riverso2013plug}). Moreover, unlike methods that rely on communication and distributed optimisation (e.g. see \cite{conte2016distributed}), the proposed method requires minimal communication and safeguards the privacy of subsystems.

In Section II, the model of the considered class of systems is presented. In Section III, the optimization problem solved by each subsystem to find the corresponding stabilizing controller is introduced. In Section IV, the proposed controller is evaluated by applying it to DC microgrids. Finally, concluding remarks are given in Section IV.

\section{Problem Formulation}

We consider discrete-time large-scale systems which can be decomposed into a set of $M$ subsystems described using the linear time-invariant (LTI) dynamics,
\begin{equation}
\label{sec2_Ldyn}
\begin{aligned}
x_i^+ &= A_i x_i + B_i u_i + F_i  v_i, \quad y_i=C_ix_i, \\
v_i &= \sum_{j \in \pazocal{N}_i^-} l_{ij}(y_j-y_i), \\
\end{aligned}
\end{equation}
where $x_i \in \mathbb{R}^{n_i}$, $u_i \in \mathbb{R}^{m_i}$ and $y_i \in \mathbb{R}^{m_i}$ are the state, input and output vectors of the $i^{th}$ subsystem respectively. For each subsystem, the set $\pazocal{N}_i^-$ is the in-neighbour set, defined as the set of subsystems whose outputs affect the subsystems's dynamics. The matrices $A_i \in \mathbb{R}^{n_i \times n_i}$, $B_i \in \mathbb{R}^{n_i \times m_i}$, $F_i \in \mathbb{R}^{n_i \times m_i}$ and $C_i \in \mathbb{R}^{m_i \times n_i}$ and the scalars $l_{ij}$ are assumed to be known. We also assume that each subsystem is controllable. Note that we consider the case in which the dimension of the output vectors of all subsystems is the same. Defining the global state vector $x=[x_1^\top,...,x_M^\top]^\top \in \mathbb{R}^{n}$, the global input vector $u=[u_1^\top,...,u_M^\top]^\top \in \mathbb{R}^{m}$ and the global output vector $y=[y_1^\top,...,y_M^\top]^\top \in \mathbb{R}^{m}$, the overall system dynamics can be written as
\begin{equation}
\label{sec2_Gdyn}
\begin{aligned}
x^+=Ax+Bu, \quad
y=Cx,
\end{aligned}
\end{equation}
where the matrices $A \in \mathbb{R}^{n \times n}$, $B \in \mathbb{R}^{n \times m}$ and $C \in \mathbb{R}^{m \times n}$ are obtained from the matrices in \eqref{sec2_Ldyn} in the obvious way.

The interconnection between subsystems can be represented by the graph $\pazocal{G}(\pazocal{V},\pazocal{E},\pazocal{W})$ where $\pazocal{V}=\{1,...,M\}$, $\pazocal{E} \subseteq (\pazocal{V} \times \pazocal{V})$ and $\pazocal{W}=\{l_{ij} \in \mathbb{R}, (i,j) \in \pazocal{E}\}$ are the set of nodes, edges and weights of the graph $\pazocal{G}$. Each node in the graph represents a subsystem. An edge exists from the $i^{\text{th}}$ node to the $j^{\text{th}}$ node if the outputs of the $i^{\text{th}}$ subsystem affect the dynamics of the $j^{\text{th}}$ subsystem.  The weight $l_{ij}$ of this edge depends on the system parameters and indicates the strength of the coupling. For each node, the sets $\pazocal{N}_i^+=\{j \in \pazocal{V}:(i,j)\in\pazocal{E}\}$, $\pazocal{N}_i^-=\{j \in \pazocal{V}:(j,i)\in\pazocal{E}\}$ and $\pazocal{N}_i=\pazocal{N}_i^+ \cup \pazocal{N}_i^-$ define the out-neighbour, in-neighbour and neighbour sets respectively. The subsystem's out-neighour set includes the subsystems  whose dynamics are affected by outputs of this subsystem.

The Laplacian matrix $L \in \mathbb{R}^{M \times M}$ of the graph $\pazocal{G}$ describes the coupling structure between the subsystems and its entries are defined as
\begin{equation}
L_{ij} = 
\left\{
\begin{matrix}
\sum_{j \in \pazocal{N}_i} l_{ij}, & i=j, \\
-l_{ij}, & i \neq j, \ j \in \pazocal{N}_i^-, \\
0, & i \neq j, \ j \notin \pazocal{N}_i^-. \\
\end{matrix}
\right.
\end{equation}

The aim of this work is to synthesize a decentralized passivity-based control law,
\begin{equation}
\label{sec2_cl}
u_i = K_i x_i, \ \forall i \in \{1,...,M\},
\end{equation}
where the control inputs of each subsystem depends on the states of the subsystem only to ensure asymptotic stability of the whole system. We also aim to synthesize this controller in a decentralized fashion. To this end, we recall the following definition.
\begin{definition}[\hspace{1sp}\cite{aliyu2017nonlinear}]
	\label{def1}
	The discrete-time system \eqref{sec2_Gdyn} is strictly passive with respect to the input-output pair $(u,y)$ if there exist a continuous storage function $V:\mathbb{R}^n \rightarrow \mathbb{R}_{\geq 0}$ with $V(0)=0$ and a dissipation rate $\gamma:\mathbb{R}^n \rightarrow \mathbb{R}_{>0}$ with $\gamma(0)=0$ such that 
	\begin{equation}
	V(x^+) - V(x) \leq y^\top u - \gamma(x).
	\end{equation}
\end{definition}
It is known that discrete time passivity generally requires feed-forward directly linking the input to the output of the system (a non-zero “$D$” matrix in linear systems \cite{aliyu2017nonlinear}, or more generally zero relative degree \cite{navarro2005several}). We note that such terms are not present in \eqref{sec2_Ldyn}. We address this difficulty below through the introduction of virtual output variables.

\section{Control Synthesis}

In this section, we synthesize the local control laws \eqref{sec2_cl} which stabilize the whole system \eqref{sec2_Gdyn} in a decentralized fashion. For this purpose, we define for each subsystem the local virtual output
\begin{equation}
z_i = y_i + D_i v_i = C_i x_i + D_i v_i,
\end{equation}
where $D_i \in \mathbb{R}^{m \times m}$ is a decision variable. \color{black} The control synthesis is carried out by solving for each subsystem a semidefinite program which guarantees that
\begin{enumerate}
	\item[(I)] each local controller \eqref{sec2_cl} passivates the corresponding subsystem \eqref{sec2_Ldyn} with respect to the local input-output pair $(v_i,z_i)$.
	\item[(II)] the local passivity of all subsystems implies the asymptotic stability of the overall system, that is, asymptotic stability is achieved if each control input $u_i$ passivates the corresponding subsystem.
\end{enumerate}
Note that the stability of the overall system (and not the stability of individual subsystems) is considered. This is because the coupling terms might destabilize the overall network even if each subsystem is asymptotically stable in the absence of coupling. First, we derive a matrix inequality for each subsystem which ensures (I) in the following lemma. 
The matrices \eqref{sec3_pLMI}, \eqref{sec3_p1_pc2}, \eqref{sec3_p1_pc3} and \eqref{sec3_p1_pc4} are given in subsequent pages in single column.

\begin{lemma}
	\label{lemma1}
	The $i^{\text{th}}$ subsystem \eqref{sec2_Ldyn} is strictly passive with respect to the input-output pair $(v_i,z_i)$ under the control law \eqref{sec2_cl} if there exist matrices $S_i \in \mathbb{R}^{m_i \times m_i}$ and $G_i \in \mathbb{R}^{m_i \times n_i}$ and positive definite matrices $E_i \in \mathbb{R}^{n_i \times n_i}$ and $H_i \in \mathbb{R}^{n_i \times n_i}$ such that the matrix inequality \eqref{sec3_pLMI} holds.
	\begin{table*}
		\normalsize
		\begin{equation}
		\label{sec3_pLMI}
		\begin{bmatrix}
		E_i & \frac{1}{2} E_i C_i^\top & (A_i E_i+B_i G_i)^\top & E_i \\
		\frac{1}{2} C_i E_i & \frac{1}{2} S_i + \frac{1}{2} S_i^\top & F_i^\top & 0 \\
		(A_i E_i + B_i G_i) & F_i & E_i & 0 \\
		E_i & 0 & 0 & H_i
		\end{bmatrix}
		\geq 0
		\end{equation}
	\end{table*}
\end{lemma}
\begin{proof}
	The closed loop dynamics of the $i^{\text{th}}$ subsystem under the controller $u_i=K_i x_i$ is given by
	\begin{equation}
	\label{sec3_p1_cld}
	\begin{aligned}
	x_i^+=(A_i+B_iK_i)x_i+F_iv_i, \quad z_i=y_i+D_iv_i.
	\end{aligned}
	\end{equation}
	According to Definition \ref{def1}, the $i^{\text{th}}$ subsystem under the controller $u_i=K_i x_i$ is strictly passive with respect to the input-output pair $(v_i,z_i)$ if and only if there exists a positive semidefinite storage function $V_i(x_i)$ and a positive definite dissipation rate $\gamma_i(x_i) > 0$ such that
	\begin{equation}
	\label{sec3_p1_pc}
	V_i(x_i^+)-V_i(x_i) \leq v_i^\top z_i - \gamma_i(x_i).
	\end{equation}
	Considering the positive definite quadratic functions $V_i(x_i)=x_i^\top P_i x_i$ and $\gamma_i(x_i)=x_i^\top \Gamma_i x_i$ and substituting \eqref{sec3_p1_cld} in \eqref{sec3_p1_pc} yield
	\begin{equation}
	\label{sec3_p1_pc1}
	\begin{aligned}
	&x_i^\top \left(P_i-(A_i+B_i K_i)^\top P_i (A_i+B_i K_i)-\Gamma_i \right) x_i \\
	& \quad \quad \quad \quad
	+ 2 v_i^\top \left(\frac{1}{2}C_i-F_i^\top P_i (A_i +B_i K_i) \right) x_i \\
	& \quad \quad \quad \quad \quad \quad \quad \quad
	+ v_i^\top \left( D_i - F_i^\top P_i F_i \right) v_i \geq 0. \\
	\end{aligned}
	\end{equation}
	Note that $v_i^\top D_i v_i = v_i^\top \left( \frac{D_i+D_i^\top}{2}+\frac{D_i-D_i^\top}{2} \right) v_i =  v_i^\top \left( \frac{D_i+D_i^\top}{2} \right) v_i$ since $\frac{D_i+D_i^\top}{2}$ is symmetric whereas $\frac{D_i-D_i^\top}{2}$ is skew symmetric. Hence, \eqref{sec3_p1_pc2} is implied by \eqref{sec3_p1_pc1}.
	\color{black}
	\begin{table*}
		\normalsize
		\begin{equation}
		\label{sec3_p1_pc2}
		\begin{bmatrix}
		P_i-(A_i+B_iK_i)^\top P_i (A_i+B_iK_i)-\Gamma_i & \frac{1}{2} C_i^\top-(A_i+B_iK_i)^\top P_i F_i \\
		\frac{1}{2}C_i-F_i^\top P_i (A_i+B_i K_i)^\top & \frac{1}{2} D_i+\frac{1}{2}D_i^\top-F_i^\top P_i F_i
		\end{bmatrix} \geq 0
		\end{equation}
		
		\begin{equation}
		\label{sec3_p1_pc3}
		\begin{bmatrix}
		P_i^{-1} - P_i^{-1} \Gamma_i P_i^{-1} & \frac{1}{2} P_i^{-1} C_i^\top \\
		\frac{1}{2} C_i P_i^{-1} & \frac{1}{2} D_i+\frac{1}{2}D_i^\top
		\end{bmatrix}
		-
		\begin{bmatrix}
		(A_i P_i^{-1} + B_i K_i P_i^{-1})^\top \\ F_i^\top \\
		\end{bmatrix}
		P_i
		\begin{bmatrix}
		(A_i P_i^{-1} + B_i K_i P_i^{-1}) & F_i^\top \\
		\end{bmatrix}
		\geq 0
		\end{equation}
		
		\begin{equation}
		\label{sec3_p1_pc4}
		\begin{bmatrix}
		P_i^{-1} & \frac{1}{2} P_i^{-1} C_i^\top &  (A_i P_i^{-1} + B_i K_i P_i^{-1})^\top \\
		\frac{1}{2} C_i P_i^{-1} & \frac{1}{2} D_i+\frac{1}{2}D_i^\top & F_i^\top \\
		(A_i P_i^{-1} + B_i K_i P_i^{-1}) & F_i^\top & P_i^{-1}
		\end{bmatrix}
		-
		\begin{bmatrix}
		P_i^{-1} \\ 0 \\ 0 \\
		\end{bmatrix}
		\Gamma_i
		\begin{bmatrix}
		P_i^{-1} & 0 & 0 \\
		\end{bmatrix}
		\geq 0
		\end{equation}
		\hrule
	\end{table*} 
	Multiplying \eqref{sec3_p1_pc2} by $\operatorname{diag}(P_i^{-1},I_{m_i})$	from both sides where $I_{m_i}$ is an identity matrix of size $m_i$ and rearranging the resulting inequality yield \eqref{sec3_p1_pc3}. Note that multiplying by $\operatorname{diag}(P_i^{-1},I_m)$ is valid since $P_i$ is positive definite. 
	Applying Schur complement to \eqref{sec3_p1_pc3} and rearranging yield \eqref{sec3_p1_pc4}. Applying Schur complement to \eqref{sec3_p1_pc4} and defining the map
	\begin{equation}
	\label{sec3_map}
	E_i = P_i^{-1}, \quad G_i = K_i P_i^{-1}, \quad H_i=\Gamma_i^{-1}, \quad S_i=D_i,
	\end{equation}
	leads to \eqref{sec3_pLMI}.
\end{proof}

Note that, under some assumptions, \eqref{sec3_p1_pc2} is equivalent to the matrix inequality mentioned in \cite{kottenstette2014relationships} which ensures passivity of discrete-time systems. The map \eqref{sec3_map} is bijective as long as $P_i$ and $\Gamma_i$ are nonsingular. These two conditions are satisfied by assumption in Lemma \ref{lemma1}. Although the matrix inequality \eqref{sec3_pLMI} is not linear with respect to the variables $P_i$, $K_i$, $\Gamma_i$ and $D_i$, it becomes linear with respect to the newly-defined variables $E_i$, $G_i$, $H_i$ and $S_i$. 

Although Definition \ref{def1} requires a positive semidefinite storage function $V_i(x_i)=x_i^\top P_i x_i$, a positive definite matrix $P_i$ is used for three reasons; to be able to multiply \eqref{sec3_p1_pc2} by $\operatorname{diag}(P_i^{-1},I_{m_i})$, to define the bijective map \eqref{sec3_map} and because the matrices $P_i$ are used later to define the Lyapunov function of the system.	Note that (11) demonstrates why passivity of the $i^{\text{th}}$ subsystem with respect to the actual output $y_i$ is not possible. If  $D_i=0$, the matrix inequality can only be satisfied if $F_i=0$ and $C_i=0$, that is only if the subsystems are decoupled. This motivates the introduction of the virtual output $z_i$ above.

To ensure stability of the interconnected system under passivity with respect to the virtual input, we introduce the following lemma. In the sequel, we define $\Gamma=\operatorname{diag}(\Gamma_1,...,\Gamma_M)$ and $D=\operatorname{diag}(D_1,...,D_M)$.

\begin{lemma}
	\label{lemma2}
	Assume that the $i^{\text{th}}$ subsystem is strictly passive with respect to the input-output pair $(v_i,z_i)$ under the controller $u_i(x_i)=K_ix_i$ for all $i \in \{1,...,M\}$. The closed-loop dynamics \eqref{sec2_Gdyn} of the global system is asymptotically stable if there exists a positive definite matrix $D$ such that
	\begin{equation}
	\label{sec3_p2_sLMIm}
	\begin{bmatrix}
	\Gamma-\epsilon_0I_n+C^\top \tilde{L} C & C^\top \tilde{L}^\top \\
	\tilde{L}C & \left(\frac{D+D^\top}{2}\right)^{-1} \\
	\end{bmatrix} \geq 0,
	\end{equation}
	where $I_n$ is an identity matrix of size $n$ and $\epsilon_0$ is a positive scalar.
\end{lemma}
\begin{proof}
	The strict passivity of the $i^{\text{th}}$ subsystem with respect to the input-output pair $(v_i,z_i)$ implies that
	\begin{equation}
	\label{sec3_p2_pc}
	V_i(x_i^+)-V_i(x_i) \leq z_i^\top v_i - \gamma_i(x_i).
	\end{equation}
	Defining the Lyapunov function $V(x)=\sum_{i=1}^M V_i(x_i)=x^\top P x$ where $P=\operatorname{diag}(P_1,...,P_M)$ and summing up \eqref{sec3_p2_pc} for all subsystems lead to
	$
	V(x^+)-V(x) = \sum_{i=1}^{M} V_i(x_i^+) - \sum_{i=1}^{M} V_i(x_i) \leq \sum_{i=1}^{M} z_i^\top v_i - \sum_{i=1}^{M} \gamma_i(x_i).
	$
	Defining the function $\gamma(x)=\sum_{i=1}^M \gamma_i(x_i)=x^\top \Gamma x$ and the vectors $z=[z_1^\top,...,z_M^\top]^\top$ and $v=[v_1^\top,...,v_M^\top]^\top$ leads to
	$ V(x^+)-V(x) \leq z^\top v - x^\top \Gamma x $.
	Recall that $z_i=C_ix_i+D_iv_i$ and $v_i=\sum_{j \in \pazocal{N}_i} l_{ij}(C_j x_j-C_i x_i)$. Consequently, $z=Cx+Dv$ and $v=-\tilde{L}Cx$ where $\tilde{L} \in \mathbb{R}^{m \times m}$ consists of the submatrices $\tilde{L}_{ij}={l}_{ij} I_{m_i} \in \mathbb{R}^{m_i \times m_i}$. Thus,
	$
	V(x^+)-V(x) \leq -x^\top (\Gamma + C^\top \tilde{L} C - C^\top \tilde{L}^\top D \tilde{L} C) x
	$.
	To guarantee the asymptotic stability of the closed loop dynamics, it suffices to ensure that 
	\begin{equation}
	\label{sec3_p2_sc1}
	\Gamma + C^\top \tilde{L} C - C^\top \tilde{L}^\top \left(\frac{D+D^\top}{2}\right) \tilde{L} C \geq \epsilon_0 I_{n},
	\end{equation}
	where $\frac{D+D^\top}{2}$ replaces $D$ using a similar argument as in Lemma \ref{lemma1}. Since $D>0$ by assumption, Schur Complement is applicable to \eqref{sec3_p2_sc1} and yields \eqref{sec3_p2_sLMIm}.
\end{proof}

The matrix $D_i$ appears in the diagonal terms in \eqref{sec3_pLMI}. Thus, the higher the eigenvalues of $D_i$, the more likely the system is passive. On the other hand, $D^{-1}$ appears in the diagonal terms in \eqref{sec3_p2_sLMI}. Thus, the higher the eigenvalues of $D_i$ are, the less likely that local passivity implies asymptotic stability. Overall, the feed-forward decision variable $D_i$ encodes a trade-off between local passivity and global stability and can be chosen neither arbitrarily large nor arbitrarily small.

Next, we note that \eqref{sec3_p2_sLMIm} is nonlinear in $\Gamma$ and $D$ and the newly-defined variables in \eqref{sec3_map} leading to a nonconvex optimization problem. Moreover, \eqref{sec3_p2_sLMIm} couples all the subsystems because of the presence of the Laplacian matrix $L$ in the off-diagonal terms. Thus, if this inequality is utilized, it has to be incorporated in the optimization problems of all subsystems implying that the synthesis is no longer decentralised. 

To address these difficulties, we define the matrices $U=\tilde{L}C \in \mathbb{R}^{m \times n}$, $W=C^\top \tilde{L}^\top \in \mathbb{R}^{n \times m}$, $U_i \in \mathbb{R}^{m_i \times n}$ and $W_i \in \mathbb{R}^{n_i \times m}$ such that $U=[U_1^\top,...,U_M^\top]^\top$ and $W=[W_1^\top,...,W_M^\top]$. In the sequel, we denote the diagonal element in the $j^{\text{th}}$ row and the $j^{\text{th}}$ column of a matrix $T_i$ by $[T_i]_j$ and the 1-norm of the $j^{\text{th}}$ row by $|T_i|_j$.

\begin{theorem}
	\label{theorem1}
	The local control laws \eqref{sec2_cl} stabilize the global system \eqref{sec2_Gdyn} if for each subsystem the following constraints are feasible,
	\begin{equation}
	\label{sec4_opt}
	\begin{aligned}
	&E_i \geq \epsilon_i I_{n_i}, \ H_i \in \pazocal{D}_+, \ S_i \in \pazocal{D}_+, \eqref{sec3_pLMI}, \\
	&[H_i]_j \leq \frac{1}{|W_i|_j+\epsilon_0}, \ \forall j \in \{1,...,n_i\}, \\
	&[S_i]_k \leq \frac{1}{|U_i|_k}, \ \forall k \in \{1,...,m_i\} \text{ s.t. } |U_i|_k>0. \\
	\end{aligned} 
	\end{equation}
	where $\pazocal{D}_+$ is the set of positive-definite diagonal matrices and $\epsilon_i \ \text{ for all } i \in \{1,...,M\}$ are positive scalars.
	
\end{theorem}

\begin{proof}
	Based on the map \eqref{sec3_map}, the positive definiteness of the matrices $P_i$ and $\Gamma_i$ is guaranteed because of the constraints $E_i \geq \epsilon_i I_{n_i}$ and $H_i \in \pazocal{D}_+$. Thus, the passivity of every subsystem is ensured under the corresponding controller in \eqref{sec2_cl} using \eqref{sec3_pLMI} as indicated by Lemma \ref{lemma1}.
	
	By definition, $\Gamma_i \in \pazocal{D}_+$ and $D_i \in \pazocal{D}_+$ since $H_i \in \pazocal{D}_+$ and $S_i \in \pazocal{D}_+$. Thus, for all $j \in \{1,...,n_i\}$ and $k \in \{1,...,m_i\}$, $[\Gamma_i]_j > 0$ and $[D_i]_k > 0$ are invertible. Note also that $\left[\left(\frac{D_i+D_i^\top}{2}\right)^{-1}\right]_j = [S_i^{-1}]_j \geq |U_i|_j$ for all $j \in \{1,...,m_i\}$ s.t. $|U_i|_j>0$ since $[S_i]_j \leq \frac{1}{|U_i|_j}$ and $D_i \in \pazocal{D}_+$. Similarly, $[\Gamma_i]_j - \epsilon_0 = [H_i^{-1}]_j - \epsilon_0 \geq |W_i|_j$ for all $j \in \{1,...,n_i\}$ since $[H_i]_j \leq \frac{1}{|W_i|_j+\epsilon_0}$. Consequently, considering the definitions of $U_i$ and $W_i$, the following LMI is satisfied by diagonal dominance.
	\begin{equation}
	\label{sec3_p2_sLMI}
	\begin{bmatrix}
	\Gamma - \epsilon_0 I_n & C^\top \tilde{L}^\top \\
	\tilde{L}C & \left(\frac{D+D^\top}{2}\right)^{-1} \\
	\end{bmatrix} \geq 0.
	\end{equation}
	Since the laplacian matrix $L$ is always positive semidefinite by definition, the matrix $\tilde{L}$ is also positive semidefinite and thus, \eqref{sec3_p2_sLMI} implies \eqref{sec3_p2_sLMIm}. Hence, the local passivity of all subsystems ensured by Lemma \ref{lemma1} implies the asymptotic stability of the global system by Lemma \ref{lemma2}.
\end{proof}

Note that all constraints are convex with respect to the decision variables. Moreover, there are no common variables between the constraints of any two subsystems. Each subsystem has its own variables $E_i$, $G_i$, $H_i$ and $S_i$ which are not shared with other subsystems. Thus, adding any local convex function $f_i$ as a cost leads to a convex optimisation problem that can be solved independently by each subsystem.	Indeed the cost function can be different for each subsystem, to reflect local preferences.
Note also that other alternatives which ensure passivity of discrete-time systems, such as the KYB conditions in \cite{aliyu2017nonlinear}, the matrix inequality in \cite{kottenstette2014relationships} and the matrix inequality \eqref{sec3_p1_pc4} do not yield a convex program when replacing \eqref{sec3_pLMI} in Theorem \ref{theorem1}.

To solve the semidefinite program of one subsystem, the corresponding matrices $U_i$ and $W_i$ are required. These matrices only depend on the weights $l_{ij}$ (which describe how this subsystem is affected by its in-neighbours) and $l_{ji}$ (which describe how this subsystem affects its out-neighbours)  as well as the matrices $C_i$ of this subsystem and its neighbours. Thus, the semidefinite program of each subsystem requires limited information from its neighbouring subsystems.  For many systems the physics of the underlying process imply that connections between subsystems are naturally symmetric $(l_{ij}=l_{ji})$; this is the case for DC microgirds considered below, but also for, e.g. thermal dynamics in buildings, action-reaction forces in mechanical systems, etc. In this case the Laplacian is symmetric and the information necessary for performing the decentralised synthesis is automatically available to each subsystem.

\section{Simulation Results}
\label{sec:simulation}

\begin{figure}
	\centering
	\scalebox{0.5}{\begin{tikzpicture}
\draw[blue!55!black, dash pattern= {on 8pt off 4pt}] (0,-1) rectangle (10.2,4.2); 
\huge
\draw (1,0) to[V, v_=$\;d_iV_{in_i}$] (1,3);
\draw (1,3) to[R,l=$R_i$] (3.5,3);
\draw (3.5,3) to[L,i>^=$I_i$,l=$L_i$,] (6,3);
\draw (6,0) to[C,l=$C_i$] (6,3);
\draw (6,0) -- (2,0);
\draw (2,0) -- (1,0); 
\draw (6,3) to[short,-o] (7.1,3) -- (8.6,3) to[american current source,l=$I_{l_i}$] (8.6,0) to[short,-o] (7.1,0) -- (6,0);
\draw[-triangle 45] (7.1,0.2) -- (7.1,2.8) node[right, midway] {$V_i$};
\draw (8.4,3) -- (10.2,3) to[short] (11.7,3);
\draw (8.4,0) -- (11.7,0);
\node at (12.95,1.6) [rectangle,draw,fill=blue!4!,line width=0.5mm, minimum height=5.2cm, minimum width=2cm] {\LARGE Microgrid};
\node at (5.1,4.7) [] {\textcolor{blue!55!black}{$i$-th DGU}}; 
\end{tikzpicture}}
	\caption{Electric circuit representing the averaged model of a DC/DC buck converter connected to the microgrid.}
	\label{DGU_circuit}
\end{figure}
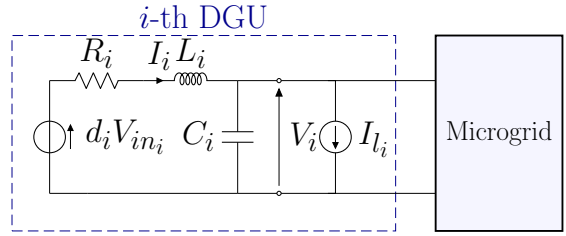
\begin{figure}
	\centering
	\scalebox{0.35}{\begin{tikzpicture}
\Huge
	\node (v1) at (0,0) [rectangle, draw, thick, fill=blue!4!] {DGU\textsubscript{1}};
	\node (v4) at (5,0) [rectangle, draw, thick, fill=blue!4!] {DGU\textsubscript{4}};
	\node (v5) at (12,0) [rectangle, draw, thick, fill=blue!4!] {DGU\textsubscript{5}};
	\node (v6) at (17,0) [rectangle, draw, thick, fill=blue!4!] {DGU\textsubscript{6}};
	\node (v2) at (5,-5) [rectangle, draw, thick, fill=blue!4!] {DGU\textsubscript{2}};
	\node (v3) at (12,-5) [rectangle, draw, thick, fill=blue!4!] {DGU\textsubscript{3}};
	
	\draw (0,-0.64) to[european resistor,l=$R_{12}$] (0,-3.5);
	\draw (5,-0.64) to[european resistor,l=$R_{24}$] (5,-3.5);
	\draw (12,-0.64) to[european resistor,l=$R_{35}$] (12,-3.5);
	\draw (17,-0.64) to[european resistor,l=$R_{36}$] (17,-3.5);
	\draw (0,-3.5) to[european resistor,l=$R_{23}$] (17,-3.5);
	\draw (5,-3.5) -- (5,-4.36);
	\draw (12,-3.5) -- (12,-4.36);
\end{tikzpicture}}
	\caption{Considered microgrid structure.}
	\label{struct}
\end{figure}
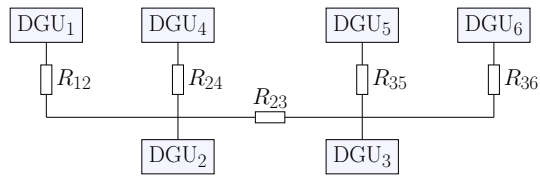
\begin{figure*}
	\centering
	\captionsetup[subfigure]{oneside,margin={0.6cm,0cm}}
	\begin{subfigure}{0.32\textwidth}
		\centering 
		\includegraphics[scale=0.33]{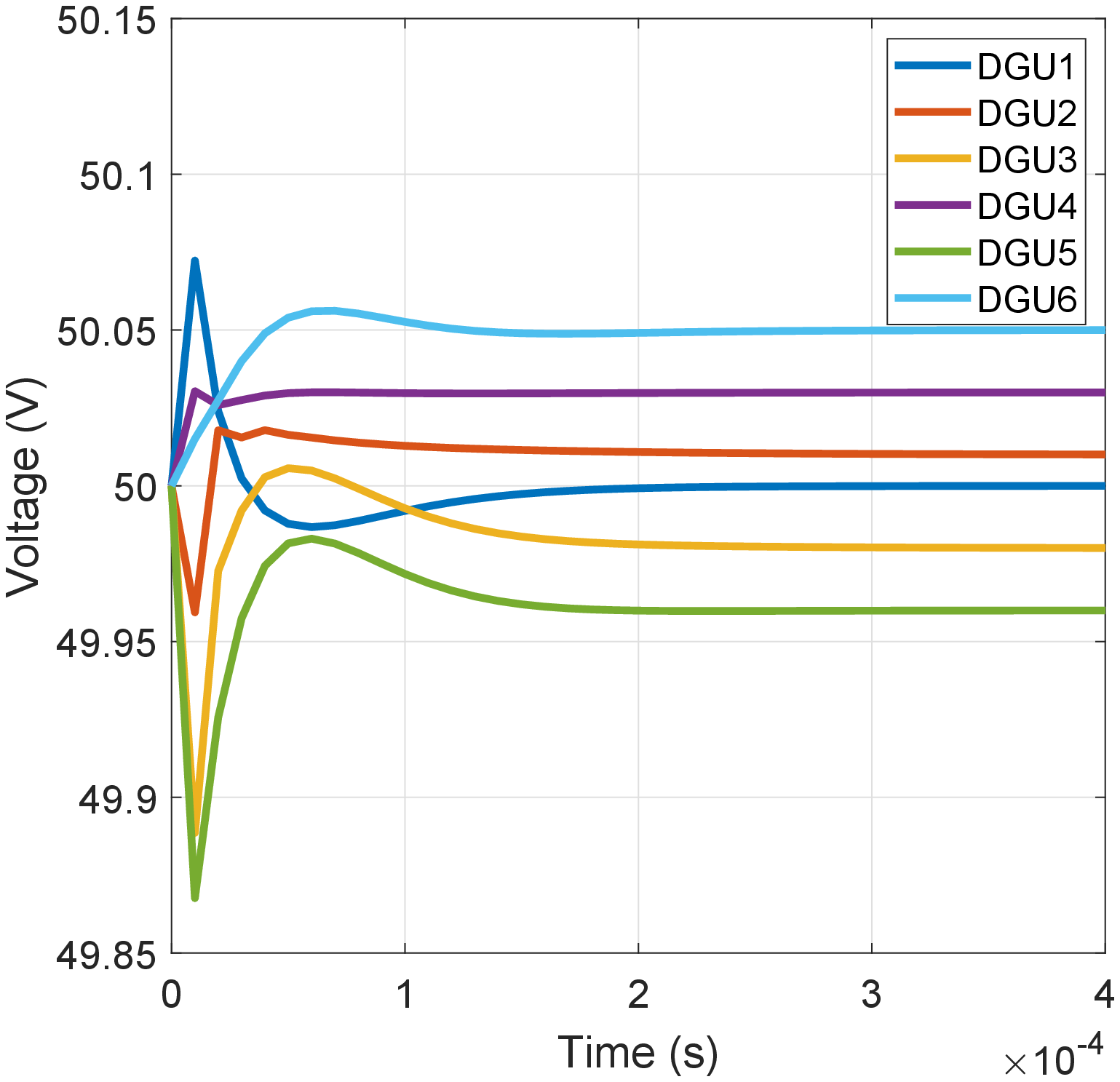}
		\label{res_fig2a}
	\end{subfigure}
	\begin{subfigure}{0.32\textwidth}
		\centering
		\includegraphics[scale=0.33]{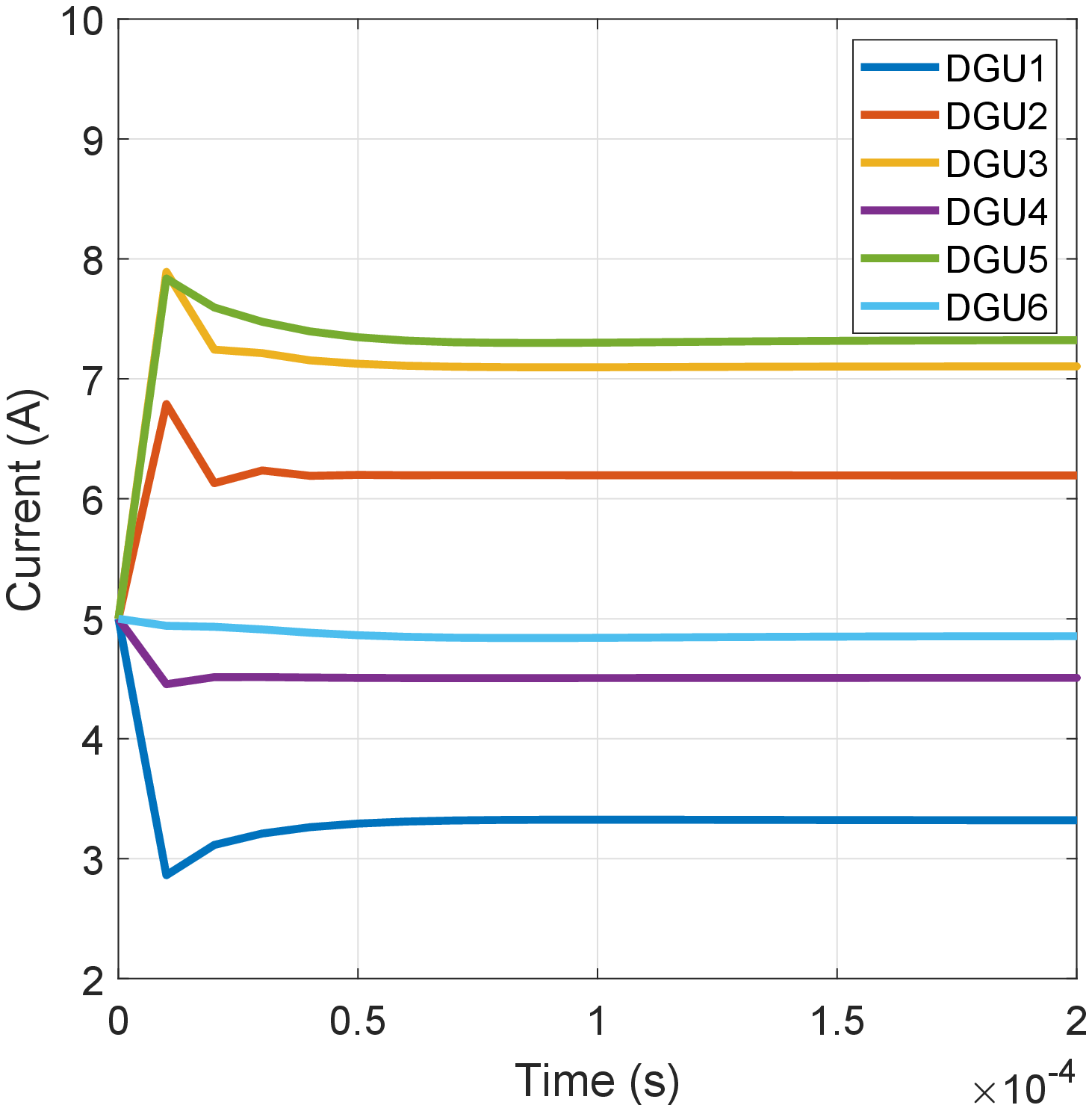}
		\label{res_fig2b}
	\end{subfigure}
	\begin{subfigure}{0.32\textwidth}
		\centering
		\includegraphics[scale=0.33]{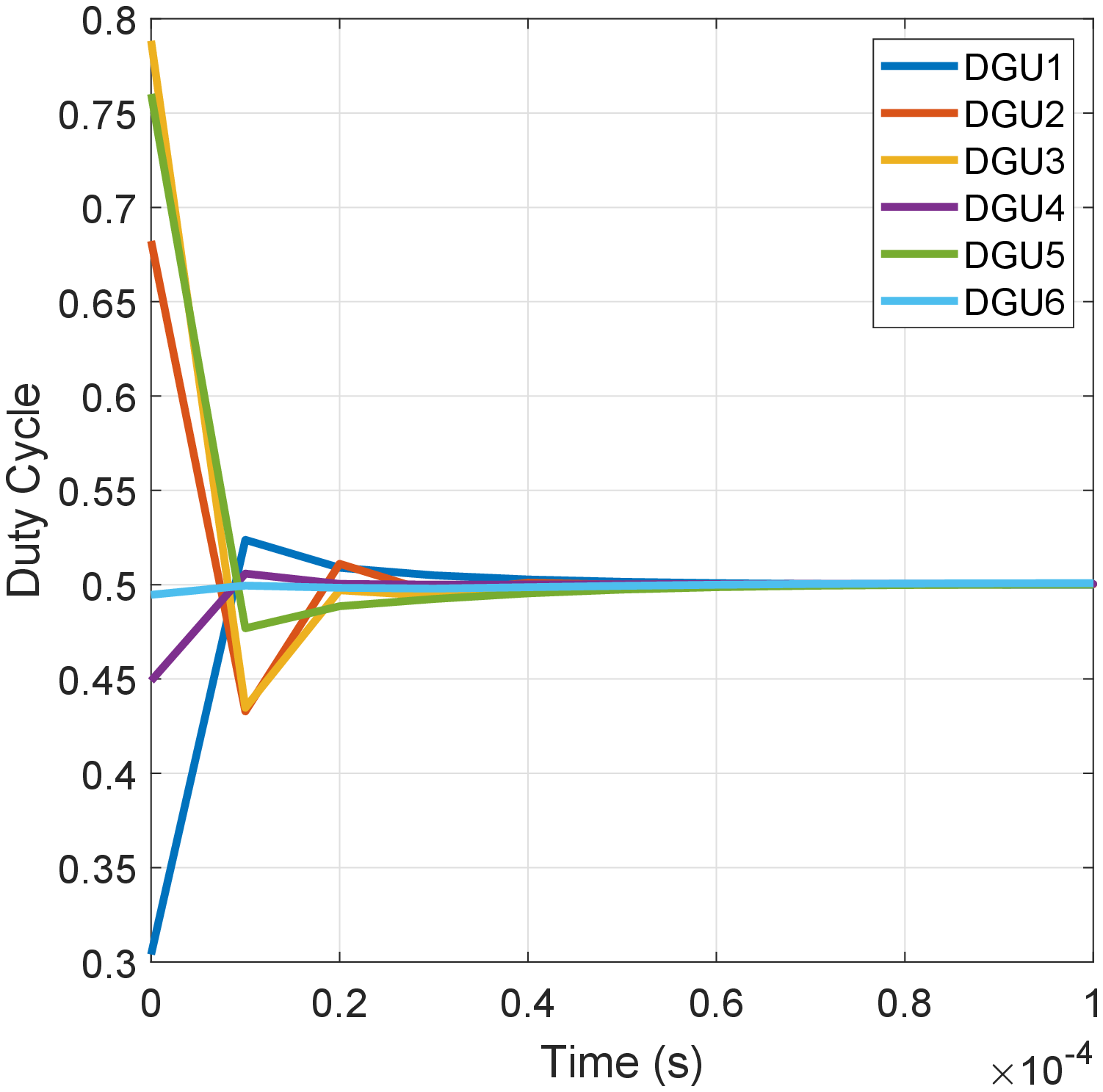}
		\label{res_fig2c}
	\end{subfigure}
	\caption{Output voltages (left), converter currents (middle) and duty cycles (right) of all DGUs when the cost $f_i^c$ is used.}
	\label{sec4_res}
\end{figure*}
We evaluate the proposed control scheme by applying it to a network of distributed generation units (DGUs). Each DGU consists of a DC voltage source and a buck converter as shown in Fig.\ref{DGU_circuit}. The voltage source represents a renewable energy source which provides a constant voltage $V_{in_i}$. The buck converter is represented by an RLC circuit with a resistance $R_i$, an inductance $L_{i}$ and a capacitance $C_i$. A switch is used to regulate the output voltage of the DGU by appropriately selecting the duty cycle $d_i$. Two neighbouring DGUs $i$ and $j$ are connected through a resitive line with a resistance of $R_{ij}$. Each DGU is assumed to support a constant current load which requires a current $I_{l_i}$.

For every DGU, let $V_i$ and $I_i$ be the output voltage and the converter current respectively. To avoid any steady state error in the output voltages, each DGU is augmented with an integrator whose state is $s_i$. Considering the state vector $x_i=[V_i,I_i-I_{l_i},s_i]^\top$ and the input vector $u_i=d_i-\frac{R_i I_{l_i}}{V_{in_i}}$, the average dynamics of the $i^{\text{th}}$ DGU can be written as
\begin{equation}
\label{sec4_Ldyn}
\begin{aligned}
\dot{x}_i &= A_{c_i} x_i + B_{c_i} u_i + F_{c_i}  v_i, \quad y_i=C_i x_i, \\
v_i &= \sum_{j \in \pazocal{N}_i} l_{ij}(y_j-y_i),
\end{aligned}
\end{equation}
where $C_i = [1 \ 0 \ 0]$, $l_{ij}=\frac{1}{R_{ij}}$,
\begin{equation*}
\begin{aligned}
A_{c_i} = 
\begin{bmatrix}
0 & \frac{1}{C_i} & 0\\
-\frac{1}{L_i} & -\frac{R_i}{L_i} & 0 \\
\alpha_i & 0 & 0 \\
\end{bmatrix},
B_i = \begin{bmatrix} 0 \\ \frac{V_{in_i}}{L_i} \\ 0 \end{bmatrix},  
F_i = \begin{bmatrix} \frac{1}{C_i} \\ 0 \\ 0 \end{bmatrix}, 
\end{aligned}
\end{equation*}
and $\alpha_i$ is the integrator coefficient. As mentioned above, DC Microgrids are represented using undirected graphs where $l_{ij}=l_{ji}$ and $\pazocal{N}_i^{-}=\pazocal{N}_i^+$.
We consider the six-DGU network given in \cite{babazadeh2018robust} whose structure is shown in Fig. \ref{struct}.

The first difficulty to be addressed is time discretisation. 
Although the microgrid model \eqref{sec4_Ldyn} and the considered model \eqref{sec2_Ldyn} have the same structure, \eqref{sec4_Ldyn} is in continuous-time whereas \eqref{sec2_Ldyn} is in discrete-time.
When applying exact discretization to \eqref{sec4_Ldyn}, the matrices of the resulting discrete-time model are dense, compromising the distributed structure. 
Recently, considerable effort has been devoted to finding discrete-time models of good accuracy that preserve the continuous-time model structure \cite{souza2015discretisation,farina2013block}. 
Here we compare four methods which preserve the model structure by computing the root mean squared error between the voltages and currents of all DGUs obtained by these methods and those obtained by exact discretization for impulsive, step and random inputs. We use a sampling time $T_s=10^{-5}s$ and select the parameter $\alpha_i=\frac{1}{T_s}$ for all DGUs.

The first (SN) and second (FN) methods compute approximate discrete-time models by solving an optimization problem which minimizes, respectively, the spectral norm and the Frobenius norm of the error between the exactly-discretized model matrices and the approximate model matrices \cite{souza2015discretisation}. Besides sampling and holding the control inputs, the third (AM) and fourth (LM) methods sample and hold, respectively,  the coupling terms $\sum_{j \in \pazocal{N}_i} \frac{1}{R_{ij}} y_j$ \cite{farina2013block} and the vector $v_i$ in \eqref{sec4_Ldyn}. Table \ref{rmse} shows that this last method leads to the highest accuracy while maintaining the desired structure; this method was therefore selected for our controller design.

\begin{table}
	\caption{The root mean squared error between the output voltages $V_i$ and converter currents $I_i$ of each model and those of the exact model in the case of an impulsive input, a step input and a random input.}
	\centering
	\normalsize
	\begin{tabular}{|c|c|c|c|c|c|c|}
		\hline
		&  SN & FN & AM & LM \\
		\hline
		Impulse & 8.45 & 3.83 & 5.06 & 0.06 \\
		\hline
		Step & 69.31 & 30.46 & 39.68 & 0.48 \\
		\hline
		Random & 34.81 & 19.2 & 21.46 & 0.33 \\
		\hline
	\end{tabular}
	\label{rmse}
\end{table}

To compute the corresponding controller, each DGU solves its local optimization problem. We solve these local problems using MATLAB with YALMIP \cite{lofberg2004yalmip} and MOSEK \cite{mosek}. Although the LM model is used in the optimization problem, the resulting controller is applied to the exactly-discretized model to evaluate its performance in simulation. We compare the proposed decentralized controller to a centralized discrete linear quadratic regulator (LQR). The LQR control gains are computed as $K_c=-(B^\top P_{c} B+R)^{-1} B^\top P_c A$ where the matrix $P_c$ is the unique positive-definite solution of the Riccati equation $P_c=A^\top P_c A + Q - A^\top  P B (B^\top P_c B + R)^{-1} B^\top P A$. 
The matrices $Q$ and $R$ are chosen to be the identity matrices $I_n$ and $I_m$ respectively.

We evaluate three different cost functions for the proposed controller. The first one $f^a_i=0$ is used to just find a feasible solution. The function $f^b_i=\operatorname{trace}(H_i)$ aims at maximizing the dissipation rate which is an indication of maximizing the passivity margin. Finally, $f^c_i=\|E_i-E_{c_i}\|_F$ tries to mimic the behaviour of the LQR by minimizing the Frobenius norm between the matrices $E_i$ and $E_{c_i}=T_i P_c^{-1} T_i^\top$  where $T_i \in \{0,1\}^{2 \times 2}$ selects the diagonal submatrix corresponding to the $i^{\text{th}}$ subsystem.

We perform 100 Monte Carlo simulations with the reference voltages changing initially from $50V$ to a random value between $49.95V$ and $50.05V$ and the load currents changing initially from $5A$ to a random value between $2.5A$ and $7.5A$. The goal is to regulate the output voltage of each DGU to the corresponding reference $V_{r_i}$ in the presence of these loads.

To converge to the desired reference, the feedforward terms $u_{f_I}=-\frac{V_{r_i}}{V_{in_i}}+K_i [-V_{r_i} \ 0 \ 0]^\top$ and $s_{f_i}=-V_{r_i}$ are added to the control input $u_i$ and the integrator state $s_i$ dynamics respectively. Although these terms lead to shifted coordinates, they change neither the system matrices nor the Laplacian matrix. Hence, neither passivity nor stability are affected since the constraints in \eqref{sec4_opt} are still satisfied. This matches the fact mentioned in \cite{jayawardhana2007passivity} that an LTI system with shifted coordinates is passive if its associated system with non-shifted coordinates is passive. Note that the control input of one DGU is a function of its local variables and parameters only (i.e. gains, states and references).

For each simulation, the tracking error magnitude 
$e=\sqrt{\sum_{k=0}^T \sum_{i=1}^6 (\Delta V_i^{k^2} + \Delta I_i^{k^2} + \Delta s_i^{k^2} + \Delta u_i^{k^2})}$ is computed where $\Delta V_i^k = V_i^k-V_{r_i}$, $\Delta I_i^k=I_i^k-I_{r_i}$, $\Delta s_i^k = s_i^k - s_{r_i}$, $\Delta u_i^k=u_i^k-u_{r_i}$, $T$ is the simulation time, $I_{r_i}$, $s_{r_i}$ and $u_{r_i}$ are the steady state values of the corresponding variables.
We denote the magnitudes of the proposed controller with the cost functions $f^a_i$, $f^b_i$ and $f^c_i$ by $e^a_{pbc}$, $e^b_{pbc}$ and $e^c_{pbc}$ respectively and that of the LQR controller by $e_{lqr}$.

The closed-loop performance of one test scenario which uses the function $f_i^c$ is given in Fig.\ref{sec4_res} that shows the output voltage $V_i$, converter current $I_i$ and duty cycle $d_i$ of all DGUs. In this scenario, the reference voltages are chosen to be $V_{r_i}=50+0.01(i-1)(-1)^i$ where $i \in \{1,...,6\}$. Despite the uncertainties due to the discretization errors, the output voltages converge to the desired reference value. This shows the inherent robustness of our approach against discretization errors.  
Note that the other cost functions resut in similar behaviours.

The mean $\mu_J^k$ and standard deviation $\sigma_J^k$ of the suboptimality indexes $J^k = \frac{e^k_{pbc}-e_{lqr}} {e_{lqr}},$ $k \in \{a,b,c\}$ are given in Table \ref{lqr}. It is found that $f_i^c$ results in a relatively good performance (i.e. small $\mu_J^c$ and $\sigma_J^c$). This could be because $f_i^c$ tries to mimic the behavior of the LQR. We conjecture that suboptimality occurs because the control gains are not exactly the same since the proposed controller is decentralized whereas LQR is centralized. On the other hand, we also conjecture that $f^b_i$ results in poor performance (i.e. large $\mu_J^b$ and $\sigma_J^b$) since it only maximizes the passivity margin. 

Table \ref{lqr} also shows the minimum eigenvalue $\underline{\lambda}^k$ of the dissipation rate matrix $\Gamma$ which indicates how strict passivity is for each cost function. This eigenvalue can be considered as a measure of robustness, for example against uncertainties due to discretization errors that may lead to loss of passivity and stability. The function $f_i^b$ results in a large eigenvalue, as opposed to $f_i^c$. Thus, we conjecture that $f_i^b$ leads to a more robust controller compared to $f_i^c$. 

When exploring the effect of the parameter $\epsilon_0$, it is found that the system is underdamped for small $\epsilon_0$ and overdamped for large $\epsilon_0$ when using $f^a_i$. In addition, larger $\epsilon_0$ leads to slower convergence with larger overshoot. On the other hand, the performance is almost the same when using $f^b_i$ and $f^c_i$. For all cost functions, the optimization problems become infeasible for very large $\epsilon_0$. The simulation results showing the effect of $\epsilon_0$ are omitted for the interest in space.

\begin{table}
	\caption{Suboptimality mean $J^k_m$, suboptimality standard deviation $J^k_s$ and minimum eigenvalue $\underline{\lambda}$ of the matrix $\Gamma$ of the proposed controller for different cost functions.}	
	\centering
	\normalsize
	\begin{tabular}{|c|c|c|c|}
		\hline
		& $f^a_i$ & $f^b_i$ & $f^c_i$ \\
		\hline
		$\mu_J$ & $0.05$ & $0.13$ & $0.02$ \\
		\hline
		$\sigma_J$ & $0.02$ & $0.02$ & $0.01$ \\
		\hline
		$\underline{\lambda}$ & $0.014$ & $0.02$ & $0.01$ \\
		\hline
	\end{tabular}
	\label{lqr}
\end{table}

\section{Conclusions}
A passivity-based control scheme is proposed for discrete-time large-scale systems, where the control synthesis and operation are decentralised. The proposed scheme ensures both passivity and stability of such systems. By appropriately choosing the cost function of the control synthesis optimization problem, the resulting controller might lead to a closed-loop behavior similar to that of LQR. Future work includes extending this approach to varying-topology networks in which various subsystems join and leave the network from time to time. The main challenge in this direction is that stability has to be ensured in the presence of changing dynamics.

\bibliographystyle{ieeetr}
\bibliography{references}

\end{document}